\documentclass[%
 reprint,
superscriptaddress,
 amsmath,
 amssymb,
 aps,
 prx,
 showpacs, 
 floatfix,
]{revtex4-2}
\usepackage{CJKutf8}

\usepackage{lipsum}
\usepackage{graphicx}

\usepackage{caption}
\usepackage{epsf,amsmath,bbold,amsfonts,stmaryrd}

\usepackage[utf8]{inputenc}
\usepackage[greek,russian,spanish,english]{babel}
\usepackage[OT2,T1,T2A,OT1]{fontenc}
\usepackage{mathrsfs}
\usepackage{appendix}
\usepackage{amssymb}
\usepackage{float}
\usepackage{color}
\usepackage{hyperref}
\hypersetup{pageanchor=false}
\usepackage{indentfirst}
\usepackage{url}
\usepackage{float}
\usepackage{caption}

\usepackage{subfigure}

\usepackage{ulem}

\begin{document}

\preprint{APS/123-QED}

\title{\textbf{QED Effects on Kerr-Newman Black Hole Shadows}}

\author{Shaobing~Yuan}
\affiliation{School of Physics, Peking University, No.5 Yiheyuan Rd, Beijing
100871, P.R. China}
\author{Changkai~Luo} \thanks{Co-first author}
\affiliation{School of Physics, Peking University, No.5 Yiheyuan Rd, Beijing
100871, P.R. China}
\author{Zezhou~Hu}
\affiliation{School of Physics, Peking University, No.5 Yiheyuan Rd, Beijing
100871, P.R. China}
\author{Zhenyu Zhang}
\affiliation{School of Physics, Peking University, No.5 Yiheyuan Rd, Beijing
100871, P.R. China}
\author{Bin Chen} \thanks{Corresponding author} \email{chenbin1@nbu.edu.cn}
\affiliation{Institute of Fundamental Physics and Quantum Technology, Ningbo University, Ningbo, Zhejiang 315211, China}
\affiliation{School of Physical Science and Technology, Ningbo University, Ningbo, Zhejiang 315211, China}
\affiliation{School of Physics, Peking University, No.5 Yiheyuan Rd, Beijing
100871, P.R. China}
\affiliation{Center for High Energy Physics, Peking University,
No.5 Yiheyuan Rd, Beijing 100871, P. R. China}

\date{\today}

\begin{abstract}

Incorporating first-order QED effects, we explore the shadows of Kerr-Newman black holes with a magnetic charge through the numerical backward ray-tracing method. Our investigation accounts for both the direct influence of the electromagnetic field on light rays and the distortion of the background spacetime metric due to QED corrections. We notice that the area of the shadow increases with the QED effect, mainly due to the fact that the photons move more slowly in the effective medium and become easier to be trapped by the black hole.

\end{abstract}

\maketitle

\section{Introduction}

Since the groundbreaking achievement of capturing the first two images of the supermassive black holes (M87* and Sgr A*, respectively) by the Event Horizon Telescope (EHT) 
\cite{EventHorizonTelescope:2019dse,EventHorizonTelescope:2019ggy,EventHorizonTelescope:2019pgp,EventHorizonTelescope:2021srq,EventHorizonTelescope:2022urf,EventHorizonTelescope:2022wkp}, the field of black hole physics has entered a new era. This achievement not only provided direct confirmation of the existence of black holes but also unveiled a trove of information about these enigmatic cosmic entities and their surrounding environments. The focal point of much research has been the black hole shadow \cite{Takahashi:2005hy,EventHorizonTelescope:2019dse}, a prominent feature of the black hole image. The precise shape of this shadow has been shown to encode critical physical parameters, such as the black hole's mass and spin \cite{Broderick:2008qf, Hioki:2009na, Kumar:2019ohr,EventHorizonTelescope:2019ggy, Tsupko:2017rdo,Bambi:2019tjh,Gralla:2020srx,Gralla:2020yvo, Tsukamoto:2014tja}. Furthermore, the study of black hole shadows has proven instrumental in addressing fundamental questions spanning a broad spectrum of topics, such as the behavior of accretion disks \cite{McKinney:2012vh, Abramowicz:2011xu, Cunha:2019hzj, Zhang:2022osx, Zhang:2024lsf}, the nature of dark matter \cite{Haroon:2018ryd, Konoplya:2019sns, Chen:2019fsq, Chen:2021lvo, Faraji:2024ein}, the dynamics of an accelerating universe \cite{Chowdhuri:2020ipb,Li:2020drn}, modified gravity theories \cite{Wei:2013kza, Grenzebach:2014fha, Cunha:2015yba, Cunha:2016wzk, Wei:2020ght, Ovgun:2020gjz, Kuang:2022ojj}, and the existence of extra dimensions \cite{Vagnozzi:2019apd, Banerjee:2019nnj}. These intriguing questions have ignited a surge of theoretical and experimental research into black hole shadows.

Theoretical investigations into black hole shadows primarily hinge on our understanding of photon trajectories in the spacetime surrounding black holes. In a vacuum, photons move along geodesics under the approximation of geometric optics. Thus, the equations of motion of photons are completely governed by the spacetime background. However, there will inevitably be other fields besides the gravitational field in real spacetime. Therefore, a realistic possible scenario is that photons interact with other fields as they travel outside the black hole. Of particular relevance are magnetic fields, which play a pivotal role in black hole physics \cite{Wald:1974np, Blandford:1977ds, Frolov:2010mi, Wang:2021ara,  Rueda:2022fgz, Hou:2023hto}. Observations have indicated the presence of strong magnetic fields around supermassive black holes, which may result in the breakdown of the superposition principle. The electromagnetic field exhibits self-interaction, or in other words, photon-photon interactions become significant. This leads to nonlinear QED corrections to electrodynamics, making the electromagnetic field appear as if it were some kind of electromagnetic medium, known as vacuum polarization. Rays of light moving within the electromagnetic field will deviate from null geodesic paths and follow different time-like trajectories depending on their polarization direction, resulting in what is known as birefringence phenomenon \cite{DeLorenci:2000yh, Novello:1999pg}. 

Nonlinear electrodynamics effects may appear in different models, among which the most famous ones are the Euler-Heisenberg model \cite{Allahyari:2019jqz, Zeng:2022pvb, Guerrero:2023unv} and the Born-Infeld-like models \cite{Okyay:2021nnh, Uniyal:2022vdu, He:2022opa, Wen:2022hkv, Guerrero:2023unv, Amaro:2023ull, Gibbons:2001sx}. The former one is a result of one-loop quantum corrections \cite{Schwinger:1951nm}, while the latter was put forward as a solution of the divergence of a point charge's self-energy \cite{Born:1934dia}. The additional nonlinear terms in electrodynamics introduce extra terms in the energy-momentum tensor, thereby indirectly affecting the trajectory of light by modifying the spacetime geometry \cite{Breton:2019arv}. It is a challenging task to construct a fully back-reacted black hole solution and study the novel features in its image. 

In this paper, we aim to explore the influence of QED effects by examining a simplified scenario involving a rotating charged black hole. Specifically, we consider a Kerr-Newman-like black hole with a magnetic charge, taking into account the linear effect of QED on the spacetime geometry.  After taking into consideration both the direct impact of the electromagnetic field on light trajectories and the distortion of the background spacetime metric caused by QED corrections, we obtain the shadow of a black hole. 

Remarkably, we notice that the area of the black hole shadow increases with the QED corrections, which is in conflict with the ones in the literature \cite{Hu:2020usx,Zhong:2021mty,Amaro:2023ull}. The discrepancy originates from a sign difference in the photon Hamiltonian. Simply speaking, the resulting photon trajectories should be timelike, while due to the wrong sign, they appear as spacelike ones in the literature. Actually, one motivation of the present work is to correct the errors in \cite{Hu:2020usx,Zhong:2021mty} written by two of the authors (Z.Z. Hu and B. Chen) with others. 

The remaining parts of the paper are structured as follows: In Section \ref{two concepts}, we provide a concise overview of the Euler-Heisenberg Lagrangian along with its resulting dispersion relations. We correct an essential sign error in the previous works \cite{Hu:2020usx,Zhong:2021mty,Amaro:2023ull}. In Section \ref{section3}, we consider the modifications to spacetime geometry due to the additional energy-momentum tensor of the electromagnetic field. In Section \ref{four shadow}, we examine the QED influence on the shadow's shape and discuss the contributions of different mechanisms of QED effects. In Section \ref{five sum} we offer a summary of our findings. Additionally, some detailed formulas and derivations are presented in Appendix \ref{append}, \ref{deviation}, and \ref{SEC}.

\section{Basics on QED birefrigence effects}\label{two concepts}

As shown in \cite{Gibbons:2001sx}, considering one-loop vacuum polarization, we get the Euler-Heisenberg effective Lagrangian for the electromagnetic field \footnote{For convenience, we adopt Gaussian units throughout the paper. Thus there will be many $4\pi$ factors appearing in comparison with previous works.}
\begin{equation}
    \mathcal{L}=\frac{1}{4\pi}\left[-\frac{1}{4}\mathcal{F}+\frac{\mu}{16\pi}\left(\mathcal{F}^2+\frac{7}{4}\mathcal{G}^2\right)\right],
\end{equation}
where variables $\mathcal{F}$ and $\mathcal{G}$ are the only two independent relativistic invariant and pseudo-invariant constructed from the Maxwell field in four dimensions
\begin{equation}\label{invariant}
    \begin{split}
        \mathcal{F}=&F^{\mu\nu}F_{\mu\nu},\\
        \mathcal{G}=&F^{\mu\nu}\left(^*F\right)_{\mu\nu}=\frac{1}{2}\epsilon^{\mu\nu\sigma\rho}F_{\mu\nu}F_{\sigma\rho}.
    \end{split}
\end{equation}
The coefficient $\mu$ reads
\begin{equation}
    \mu=\frac{2\hslash^3\alpha^2}{45 m_e^4},
\end{equation}
with $\hslash$, $\alpha$, and $m_e$ being the Planck constant, fine-structure constant, and electron mass, respectively. We take the geometric units in this paper, setting $c=G=1$. Also, we would like to set $M=1$ for simplicity. In this case, $\mu$ is not a constant anymore because the physical constants there have dimensions, e.g. $\hslash$ is proportional to $m_{p}^{2}$. The numerical value of $\mu$ in geometric units can be calculated by
\begin{equation}\label{mu}
    \mu=\frac{2\hslash^3c^3\alpha^2}{45 G^3m_e^4M^2}\approx 9\times10^7\left(\frac{M_\odot}{M}\right)^2,
\end{equation}
where $M_\odot$ represents the mass of the sun. So, in the rest of the paper, when we change the value of $\mu$, we actually change the mass of the black hole $M$, in our geometric units.

As shown in previous works,  we can derive the modified equations of motion from the effective Lagrangian using the effective metric. The accurate expression of the effective metric is rather complicated, and we leave it to Appendix \ref{append}.
At the first order of approximation in $\mu$, the invariant effective metric tensor with upper index is
\begin{equation}
\label{first-order metric}
    \Tilde{G}^{\alpha\beta}=g^{\alpha \beta} + \frac{\lambda}{4\pi} F^{\mu\alpha} F_{\mu}^{\,\, \beta} + O\left(\lambda^2\right),
\end{equation}
with
\begin{equation}\label{lambda}
  \lambda = - 8 \mu, \quad\mathrm{or} \quad- 14\mu.
\end{equation}
The choice of $\lambda$ is determined by the polarization of the photon because the one-loop vacuum polarization makes the electromagnetic field an anisotropic dielectric. Then, the effective metric tensor with lower indices is defined to be the inverse of the $\Tilde{G}^{\alpha\beta}$,
\begin{gather}
    \Tilde{G}^{\mu\nu} \Tilde{G}_{\nu\sigma}\equiv\delta^\mu_\sigma.
\end{gather}
To the leading order of $\mu$, we read 
\begin{equation}\label{effmetric}
    \Tilde{G}_{\alpha\beta}=g_{\alpha \beta} - \frac{\lambda}{4\pi} F^{\mu}_{\,\, \alpha} F_{\mu \beta}+O\left(\lambda^2\right)\,.
\end{equation}

With the dual vector $q_\mu$ being defined as $\Tilde{G}_{\mu\nu} p^\nu$, the Hamiltonian and the Hamiltonian equations read
\begin{equation}
    \begin{split}
        H\left(q_\mu,x^\mu\right)=\frac{1}{2}\Tilde{G}^{\mu\nu}q_\mu q_\nu,\\
        \Dot{x}^\mu=\frac{\partial H}{\partial q_\mu},\hspace{3ex}\Dot{q}_\mu=-\frac{\partial H}{\partial x^\mu}.
    \end{split}
\label{eom}
\end{equation}
From the above equations, one can see that the photon effectively becomes a time-like particle in the original spacetime,
\begin{equation}\label{velocity}
    g_{\mu\nu}p^\mu p^\nu=g_{\mu\nu}\Dot{x}^\mu \Dot{x}^\nu\leq 0.
\end{equation}
Intuitively, this is just reasonable: the trajectories of photons can no longer remain light-like since they are no longer free massless photons; Instead, they become time-like under the constraint of causality. Accordingly, such interactions slow the photons down and make it easier for them to get trapped by the black hole. In other words, the gravitational traps of the photons get enhanced.

Even without gravity, QED birefringence alone can trap photons, as shown in \cite{Novello:1999pg}, where this effect was referred to as ``electromagnetic traps". Intuitively, the stronger the electromagnetic field is, the larger the refractive index of the effective medium is,  the slower the photons are. As long as the strong electromagnetic field we are considering is local, i.e. the electromagnetic field attenuates to zero at infinity, the effective medium works as a convex lens, converging the light rays, even trapping them, such that the black hole shadow may get enlarged.

The deformation of the black hole shadows due to the QED effect has been discussed in the previous works \cite{Hu:2020usx,Zhong:2021mty}. However, there is a minus sign omitted in the previous works, e.g. in Eq.(2.10) of \cite{Hu:2020usx} in comparison with Eq.(\ref{effmetric}), which would effectively appear as the opposite valued $\lambda$ in the image of the result. All the theoretical results there should be corrected by changing $\lambda$ to $-\lambda$. Then the major features of the images are also opposite.

\begin{figure*}
    \centering
    \includegraphics[width=\textwidth]{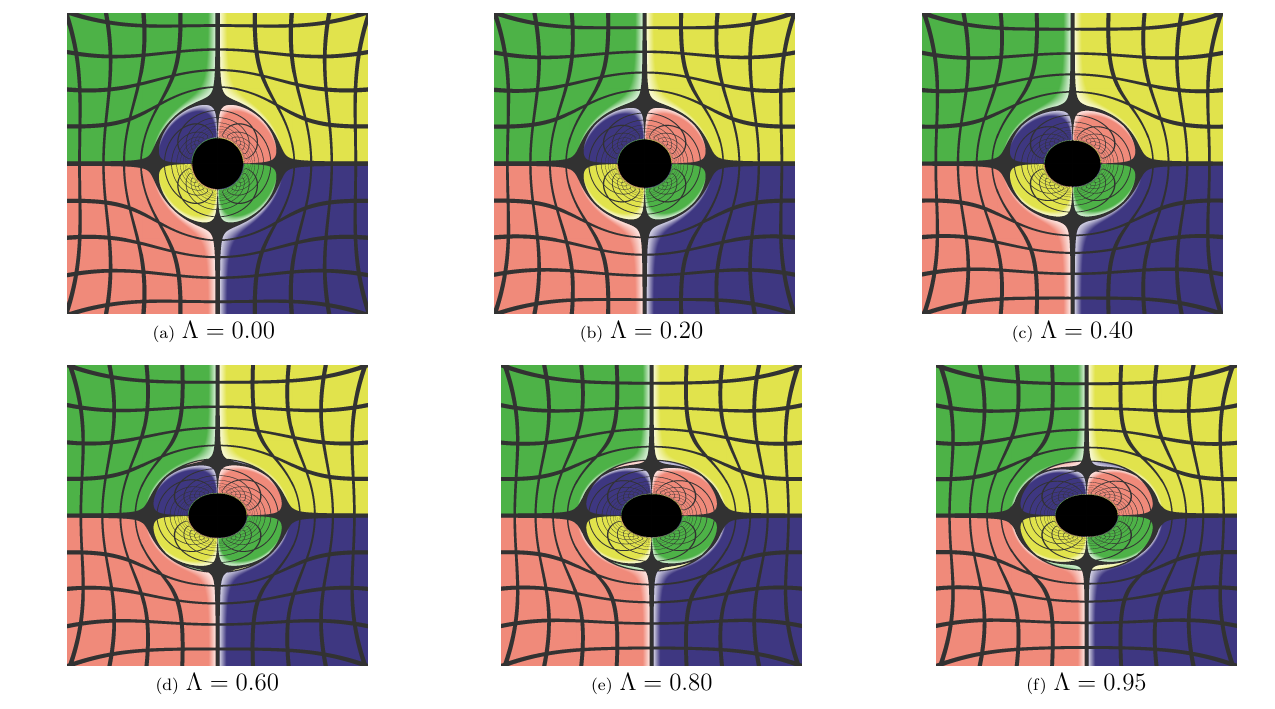}
    \caption{The images of the static black hole in uniform magnetic fields. The inclination angle of the observer is fixed at $\theta_o=\pi/2$.}
  \label{uniform1}
\end{figure*}

\begin{figure}[htbp!]
    \centering
    \includegraphics[width=3in]{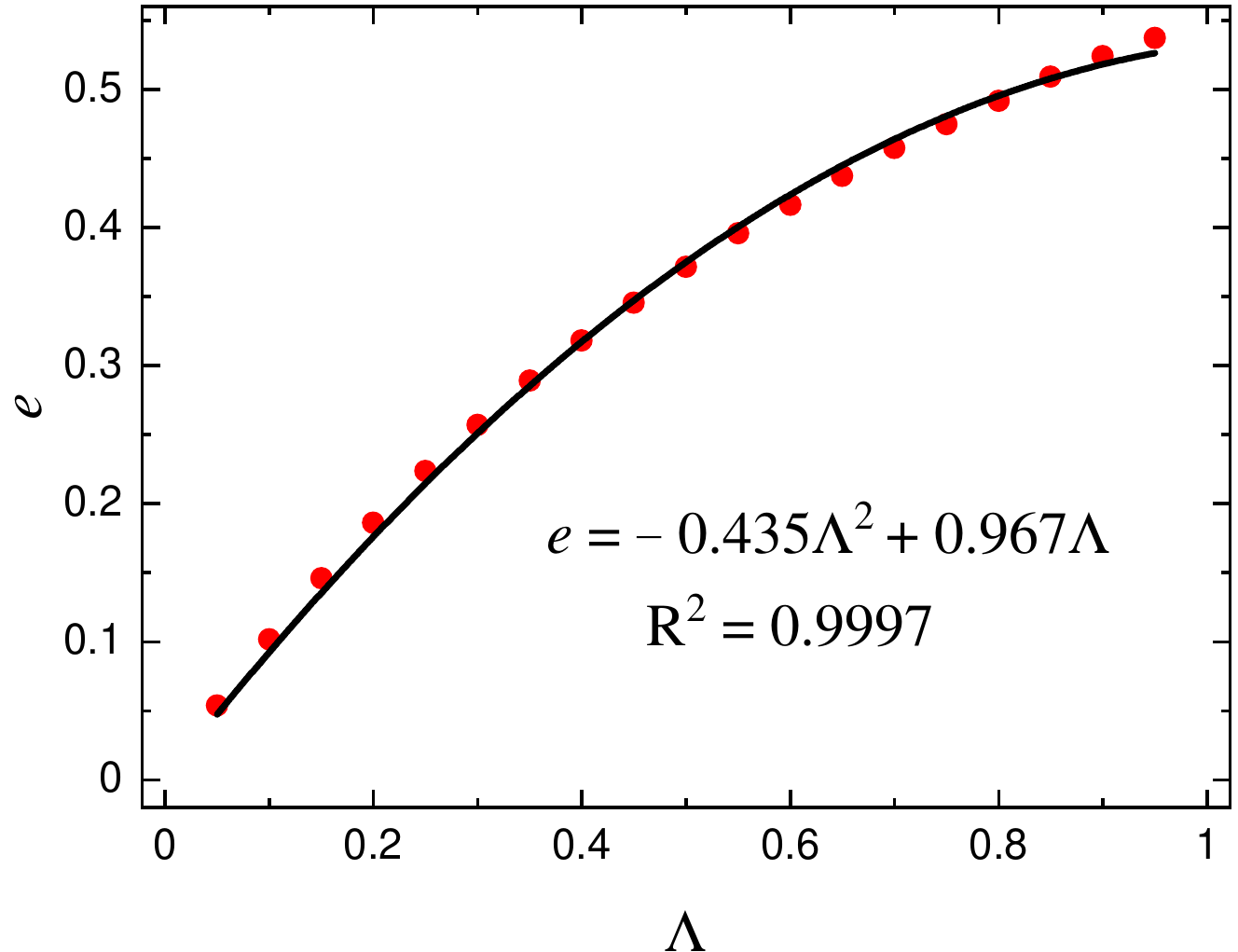}
    \caption{The variation of the eccentricity of the image concerning $\Lambda$. We use a quadratic function to fit the data.}
\label{eccentricity}
\end{figure}

\begin{figure*}
    \centering
    \includegraphics[width=\textwidth]{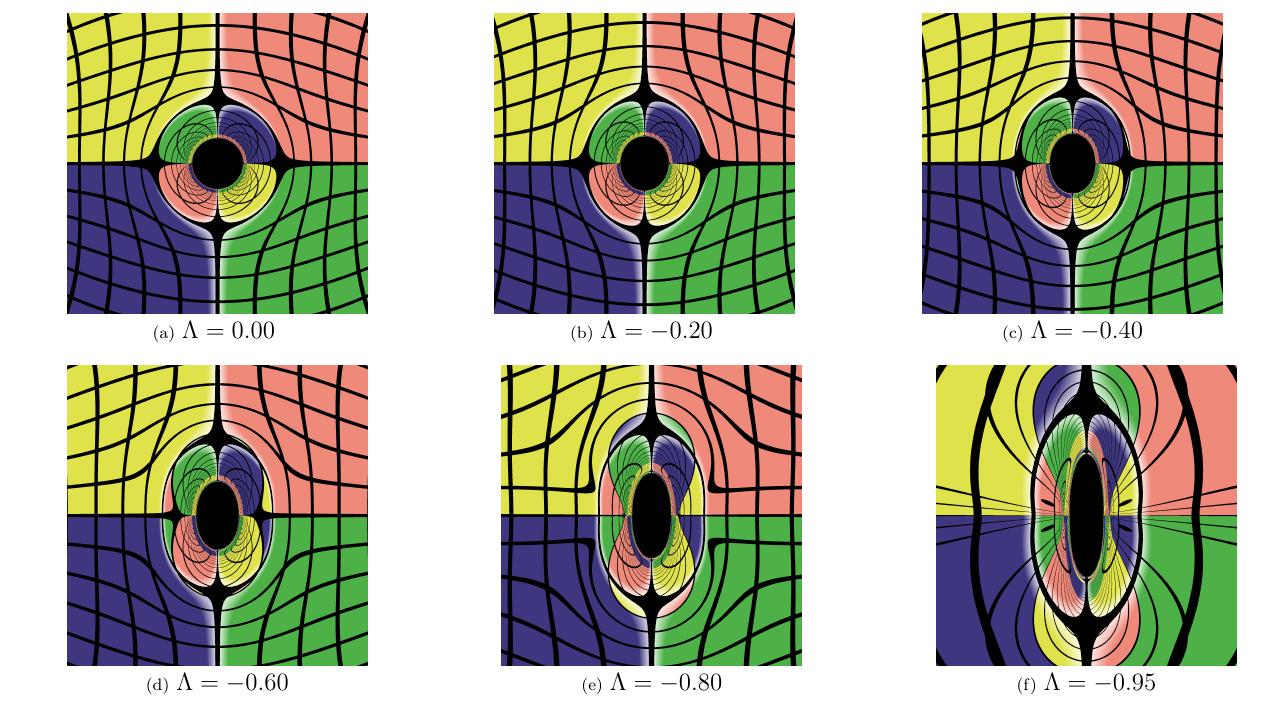}
    \caption{Fig. 3 in \cite{Hu:2020usx}.}
  \label{uniform1Pre}
\end{figure*}

\begin{figure}[htbp!]
\centering
\includegraphics[width=3in]{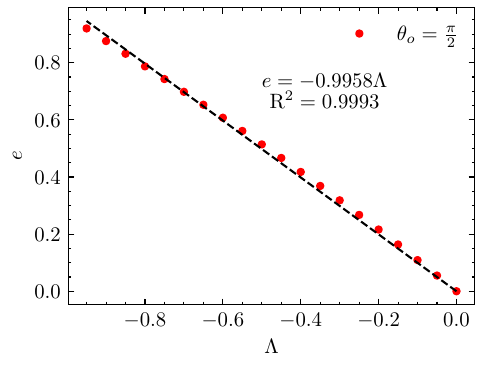}
\caption{Fig. 7 in \cite{Hu:2020usx}.}
\label{eccentricityPre}
\end{figure}

Corrected calculations of \cite{Hu:2020usx} can be briefly explained as follows. As in section 4 of \cite{Hu:2020usx}, we consider a Schwarzschild black hole immersed in a uniform magnetic field
\begin{equation}\label{uniform A}
    A=\frac{B}{2} r^2 \sin^2{\theta} \mathrm{d}\phi.
\end{equation}
The corrected effective metric is given by
\begin{equation}
    \begin{split}
        &\mathrm{d}s^2=- f (r) \mathrm{d}t^2 + \left( \frac{1}{f(r)} +\Lambda \sin^2\theta
        \right) \mathrm{d}r^2\\
        &+ \Lambda r\sin\left(2\theta\right)\mathrm{d}r\mathrm{d}\theta + r^2 \left(1 + \Lambda \cos^2\theta\right) \mathrm{d}\theta^2\\
        &+ r^2 \sin^2 \theta
        \left(1 + \Lambda \left(f (r) \sin^2\theta + \cos^2\theta\right)\right) \mathrm{d}\phi^2,
    \end{split}
\end{equation}
where $f\left(r\right)\equiv1-2/r$ is the redshift factor of an ordinary Schwarzschild black hole and the positive dimensionless quantity $\Lambda$ is defined as
\begin{equation}
    \Lambda\equiv-\frac{\lambda}{4\pi} B^2.
\end{equation}
The dimensionlessness of $\Lambda$ implies the invariance of its expression under the unit change; in other words, its expression in SI units looks the same. Consequently, the absolute strength of the magnetic field $B$ turns out to be the only quantity that affects the shape of black hole shadows in this static model while the black hole mass $M$ has no effect on it. Then the corrected images of the static black hole in uniform magnetic fields are shown in Fig. \ref{uniform1} and Fig. \ref{eccentricity}, which should be compared with Fig. \ref{uniform1Pre} and Fig. \ref{eccentricityPre} (i.e. Fig. 3 and Fig. 7 in \cite{Hu:2020usx}), respectively. Notably, the magnetic field is horizontally set in the images. As can be seen from the comparison, the shadows are stretched in the direction of the magnetic field and squeezed in the perpendicular direction rather than the other way around. Similar recalculations and discussions about results in \cite{Zhong:2021mty} can be carried out as well, which is about the rotating black holes in the magnetic fields.

We also notice that another group has conducted research in similar settings \cite{Amaro:2023ull} and got the opposite conclusion on the black hole shadow. Similar to \cite{Hu:2020usx,Zhong:2021mty}, the authors in \cite{Amaro:2023ull} made the same mistake on the sign of $\lambda$. Just like any normal GR physicists, they used the ``East Coast" metric, i.e. $\eta=\mathrm{diag}(-,+,+,+)$. However, they quoted the expression of effective metric as $\Tilde{G}^{\mu\nu}=g^{\mu\nu}+\frac{16\alpha^2\hslash^3}{45m_e^4}T^{\mu\nu}$ from \cite{osti_4071071}, which is written in the ``West Coast" metric, i.e. $\eta=\mathrm{diag}(+,-,-,-)$. Under the notation shift, the spacetime metric $g^{\mu\nu}$ shows a sign flip, while the energy-momentum tensor $T^{\mu\nu}$ does not. Therefore, the right expression in line with east coast metric should be $\Tilde{G}^{\mu\nu}=g^{\mu\nu}-\frac{16\alpha^2\hslash^3}{45m_e^4}T^{\mu\nu}$, which accords with Eq.(\ref{first-order metric}) with $\lambda=-8\mu$ up to a conformal factor \footnote{See Appendix \ref{append}}. Unfortunately, the authors in \cite{Amaro:2023ull} did not make corresponding changes and used the wrong sign of $\lambda$, which led to an opposite conclusion that black hole shadows shrink under QED effects.

\section{Back-reacted charged black hole}\label{section3}

To consider the backreaction of the QED effect, we need to study the  Einstein equation
\begin{equation}
    G_{\mu\nu}=8\pi T_{\mu\nu},
\end{equation}
with 
\begin{equation}\label{electromagnetic energy-momentum tensor}
    \begin{split}
        4\pi T_{\mu\nu}=&\left[F_{\mu}^{\,\,\alpha}F_{\nu\alpha}-\frac{1}{4}\mathcal{F}g_{\mu\nu}\right]\\
        &-\frac{\mu}{16\pi}\left[8\mathcal{F} F_{\mu}^{\,\,\alpha}F_{\nu\alpha}-\left(\mathcal{F}^2-\frac{7}{4}\mathcal{G}^2\right)g_{\mu\nu}\right].
    \end{split}
\end{equation}
For a static and spherically symmetric black hole with electric charge or magnetic charge only, its metric can be solved analytically \cite{Yajima:2000kw,Breton:2019arv}
\begin{equation}\label{QED RN}
    \begin{split}
        \mathrm{d}s^2=&-\left(1-\frac{2m\left(r\right)}{r}\right)\mathrm{d}t^2+\left(1-\frac{2m\left(r\right)}{r}\right)^{-1}\mathrm{d}r^2\\
        &+r^2\left(\mathrm{d}\theta^2+\sin^2\theta\mathrm{d}\phi^2\right),
    \end{split}
\end{equation}
where \begin{equation}
    m\left(r\right)=1-\frac{Q^2}{2r}+\frac{\mu Q^4}{20\pi r^5}
\end{equation} and $Q$ is either the electric or the magnetic charge. This black hole will be referred to as the QED-RN black hole. Notice that
\begin{equation}
    \frac{\mathrm{d}m(r)}{\mathrm{d}r}=\frac{Q^2}{2r^2}-\frac{\mu Q^2}{4\pi r^6}=4\pi r^2\left(-T_0^0\right)=4\pi r^2\rho_m.
\end{equation}
The fact that $\lim_{r\to\infty}m(r)=1=M$ indicates that $M$ actually stands for the full mass in space, including the energy of the electromagnetic field. Accordingly, with $M$ fixed, larger $Q$ means stronger electromagnetic fields. As the effective mass decreases with increasing $Q$, the shadows of classical RN as well as QED-RN black holes get smaller with increasing $Q$. However, it is remarkable that the shadow of the QED-RN black hole is larger than the one of RN black hole with the same charge $Q$, due to the fact that the QED effect screens the charge $Q^2 \to Q^2 -\mu Q^4/(10\pi r^4)$ such that the ``gravitational mass" of electromagnetic fields get reduced and the remaining gravitational mass gets larger. In other words, the QED backreaction correction tends to enlarge the shadow.

Unfortunately, when it comes to rotating cases, an exact solution has not been found yet. Nevertheless, we are going to work with the solution found in \cite{Breton:2019arv}, which is generated from the static one by using the Newman-Janis algorithm. The solution could be written in terms of the Gürses-Gürsey metric \cite{Gurses:1975vu} 
\begin{equation}
    \begin{split}
        \mathrm{d}s^2=&-\left(1-\frac{2m\left(r\right)r}{\rho^2}\right)\mathrm{d}t^2+\frac{\rho^2}{\Delta}\mathrm{d}r^2+\rho^2\mathrm{d}\theta^2\\
        &-\frac{4am\left(r\right)r\sin^2\theta}{\rho^2}\mathrm{d}t\mathrm{d}\phi+\frac{\Sigma\sin^2\theta}{\rho^2}\mathrm{d}\phi^2,
    \end{split}
\label{space-time}
\end{equation}
where
\begin{equation}
    \begin{cases}
        m\left(r\right)=1-\frac{Q^2}{2r}+\frac{\mu Q^4}{20\pi r^5},\\
        \Delta=r^2+a^2-2m\left(r\right)r,\\
        \Sigma=\left(r^2+a^2\right)^2-a^2\Delta\sin^2\theta,\\
        \rho^2=r^2+a^2\cos^2\theta.
    \end{cases}
\label{spacetime}
\end{equation}
The parameter $Q$ can be either electric charge $Q_e$ or magnetic charge $Q_m$. The outer horizon is still the largest root of $\Delta=0$, i.e.
\begin{equation}\label{hori}
10\pi r^4\left(r^2-2r+a^2+Q^2\right)=\mu Q^4,
\end{equation}
which is determined by $a$, $Q$ and $\mu$ together. Remarkably, the Newman-Janis algorithm is accurate only with linear sources. Due to the nonlinearity caused by QED effects, the above solution is only an approximation. Further discussions can be found in Appendix \ref{deviation}.

The fact that the Gürses-Gürsey metric can simply be obtained from the classical Kerr-Newman metric by using the substitute $Q^2 \to Q^2-\mu Q^4/(10\pi r^4)$ was interpreted as a screening effect in \cite{Ruffini:2013hia}, as in spherical case. Similar arguments were carried out in \cite{Amaro:2023ull}, only with a different algebraic expression describing this effective screening, which would not affect anything qualitatively.

As in the spherical case, the ``effective screening" term actually means that the QED effect reduces the ``gravitational mass" of electromagnetic fields. As we can see from Eq.(\ref{electromagnetic energy-momentum tensor}), the QED correction term in the energy-momentum tensor of electromagnetic fields reads
\begin{equation}\label{DeltaT}
    4\pi\Delta T_{\mu\nu}=-\frac{\mu}{16\pi}\left[8\mathcal{F} F_{\mu}^{\,\,\alpha}F_{\nu\alpha}-\left(\mathcal{F}^2-\frac{7}{4}\mathcal{G}^2\right)g_{\mu\nu}\right],
\end{equation}
which violates the strong energy condition (SEC) as well as the null energy condition (NEC), as shown in Appendix \ref{SEC}. More explicitly, the energy density of the QED correction term tends to be negative and reduces the gravitational energy of the electromagnetic fields. In this sense, we would expect the backreaction effect to enlarge the shadow, even at the nonlinear level. 

It was proved in \cite{Breton:2019arv} that, in the framework of nonlinear electrodynamics, the natural variables are the dual
Plebański variables ($P_{\mu\nu}$) when considering electrically charged black holes, while they are the standard Maxwell variables ($F_{\mu\nu}$) when considering magnetically charged black holes. We have not found a good method to handle the case with both charges yet. Since these two cases are similar, we mainly focus on the magnetically charged case in this work, leaving the electrically charged one for future study. For the magnetic case, we can directly get the solution of Maxwell variables \cite{Breton:2019arv}
\begin{equation}
    A_\mu\mathrm{d}x^\mu=\frac{Q_m\cos{\theta}}{\rho^2}\left[-a\mathrm{d}t+(r^2+a^2)\mathrm{d}\phi\right],
\end{equation}
\begin{equation}
    \begin{split}
        F_{\mu\nu}=-&\frac{Q_m}{\rho^4}2ar\cos\theta
        \begin{pmatrix}
        0 & 1 & 0 & 0\\
        -1 & 0 & 0 & a\sin^2\theta\\
        0 & 0 & 0 & 0\\
        0 & -a\sin^2\theta & 0 & 0
        \end{pmatrix}\\
        -&\frac{Q_m}{\rho^4}\left(r^2-a^2\cos^2\theta\right)\sin\theta\\
        &\times
        \begin{pmatrix}
        0 & 0 & a & 0\\
        0 & 0 & 0 & 0\\
        -a & 0 & 0 & \left(r^2+a^2\right)\\
        0 & 0 & -\left(r^2+a^2\right) & 0
        \end{pmatrix}.
    \end{split}
    \label{field}
\end{equation}

\section{Black hole shadows of back-reacted KN black hole}\label{four shadow}

Generally speaking, the geodesic equation in spacetime generated from static spherically symmetric metric through the Newman-Janis algorithm can be separated completely \cite{Shaikh:2019fpu}. However, we are dealing with the effective metric Eq.(\ref{effmetric}), which destroys the separability of the corresponding Hamilton-Jacobi equation. Therefore, we would like to apply the numerical backward ray-tracing method proposed in \cite{Hu:2020usx} for our study. The requisite setup includes background space-time [Eq.(\ref{space-time}) and Eq.(\ref{spacetime})], the electromagnetic field [Eq.(\ref{field})], and the photon's equation of motion [Eq.(\ref{effmetric}) and Eq.(\ref{eom})]. Here, we also employ the stereographic projection, which is often called the fisheye camera model. The details can be found in the appendix of \cite{Hu:2020usx}. Using these methods, we are able to simulate the image of the black hole with a shadow in the middle.

In our geometric units, the mass $M$ is set to 1, and alternatively the coefficient $\mu$ is variable, as explained in Section \ref{two concepts}. Also, $Q_{e}$ is set to zero for simplicity. To make the difference between the shadows with and without QED effects more significant, we choose the polarization of photons such that $\lambda=-14\mu$, i.e. $\Omega=\Omega_-$. Therefore, our three independent parameters are $a$, $Q_m$ and $\mu$. When we set $\mu=0$, we come back to the classical Kerr-Newman case.

\begin{figure}[h!]
    \centering
    
    \subfigure[$\mu=0$]{
    \begin{minipage}[t]{0.5\linewidth}
    \centering
    \includegraphics[width=1.4in]{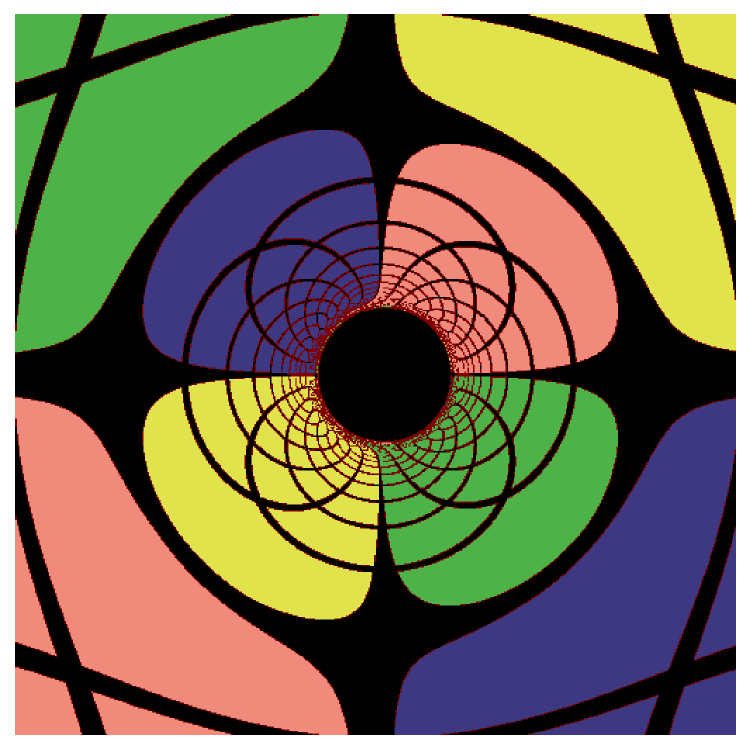}
    \end{minipage}%
    }%
    \subfigure[$\mu=1000$]{
    \begin{minipage}[t]{0.5\linewidth}
    \centering
    \includegraphics[width=1.4in]{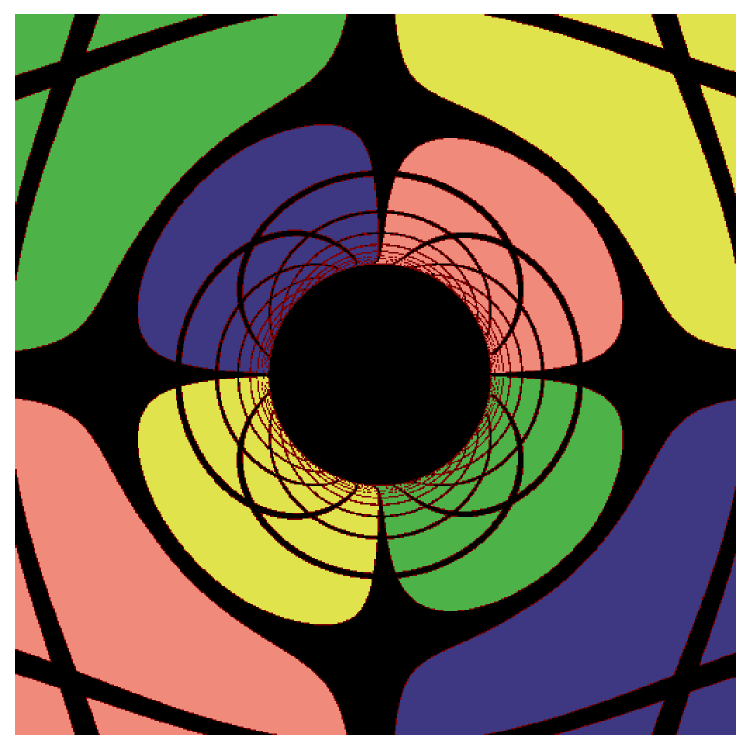}
    \end{minipage}%
    }%

    \centering
    \caption{The images of the charged spinning black hole with different QED coupling strength. The inclination angle of the observer is fixed at $\theta_o=\pi/2$, the magnetic charge is $Q_m=0.6$ and the spin is $a=0.3$.}
    \label{example}
\end{figure}

For the back-reacted KN black hole, the larger $\mu$ is, the smaller the mass is, the stronger the QED effects are and the larger the shadows are. In Fig. \ref{example}, we compare the images of the black hole with two different $\mu$: the QED effect in expanding the shadow of a black hole can be obvious. Similarly, increasing $Q_m$ leads to stronger electromagnetic fields and therefore stronger QED effects and larger expansion in shadows. In contrast, the effects of altering $a$ are rather small in a large parameter space range, as will be shown in Subsection \ref{geometrical analysis}. 

\subsection{Geometrical analysis}\label{geometrical analysis}

Roughly speaking, larger $\mu$ means stronger QED coupling, larger $Q_m$ leads to stronger electromagnetic field strength, and therefore larger $\mu$ and large $Q_m$ give rise to larger QED effects. In order to investigate their impacts on the shadows in a quantitative way, we may introduce a few geometry parameters suggested in \cite{Cunha:2015yba, Zhong:2021mty, Zhang:2022osx} as follows. The center of the shadow is defined to be 
\begin{equation}
    x_c=\frac{x_{min}+x_{max}}{2},\hspace{3ex}
    y_c=\frac{y_{min}+y_{max}}{2}=0.
\end{equation}
The polar coordinates ($\Tilde{r}$,$\alpha$) can be defined with respect to the center of the shadow as
\begin{equation}
    \begin{split}
        \Tilde{r}=\sqrt{\left(x-x_c\right)^2+y^2},\hspace{3ex}
        \tan\alpha=\frac{y}{x-x_c}.
    \end{split}
\end{equation}
Then the area of shadow can be calculated by
\begin{equation}
    S=\frac{1}{2}\int_0^{2\pi}\Tilde{r}\left(\alpha\right)^2\mathrm{d}\alpha,
\end{equation}
which can be normalized by the shadow area of the corresponding KN black hole ($\mu=0$)
\begin{equation}
    S_0=\frac{1}{2}\int_0^{2\pi}\Tilde{r}_{\text{KN}}\left(\alpha\right)^2\mathrm{d}\alpha.
\end{equation}
The ratio $S/S_0$ indicates the expansion of shadows induced by QED corrections. Additionally, the QED-induced deviation from the corresponding KN black hole ($\mu=0$) is characterized by
\begin{equation}
    \sigma_K\equiv\sqrt{\frac{1}{2\pi}\int_0^{2\pi}\left(\frac{\Tilde{r}\left(\alpha\right)-\Tilde{r}_{\text{KN}}\left(\alpha\right)}{\Tilde{r}_{\text{KN}}\left(\alpha
    \right)}\right)^2\mathrm{d}\alpha}.
\end{equation}

\begin{figure*}
    \centering
    \includegraphics[width=\textwidth]{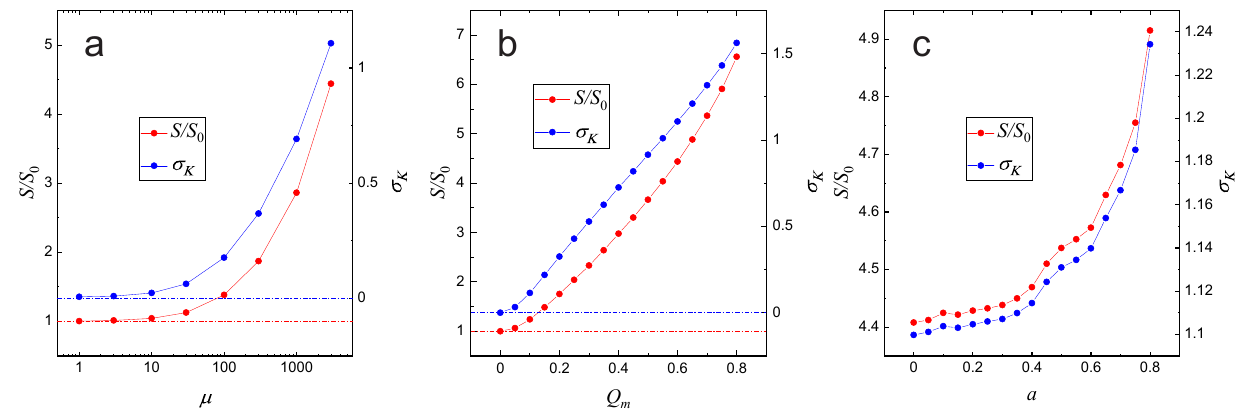}
    \caption{The variation of geometry parameters with respect to different $\mu$, $Q_m$ and $a$. $S$ refers to the area with given $\mu$ and $S_0$ refers to that with $\mu=0$. $\sigma_K$ refers to the ``normalized" deviation between given $\mu$ and $\mu=0$. \textbf{(a)}$S/S_0-\mu$ and $\sigma_K-\mu$ with $a=0.3$ and $Q_m=0.6$. \textbf{(b)}$S/S_0-Q_m$ and $\sigma_K-Q_m$ with $a=0.3$ and $\mu=3000$. \textbf{(c)}$S/S_0-a$ and $\sigma_K-a$ with $Q_m=0.6$ and $\mu=3000$.}
    \label{-Qm-a}
\end{figure*}

With the parameters defined above, we are able to discuss the strength of QED effects and its dependence on $\mu$, $Q_m$ and $a$ quantitatively. In Fig. \ref{-Qm-a}, we show how the expansion $S/S_0$ and deviation $\sigma_K$ vary with respect to different $\mu$, $Q_m$ and $a$. In the study, we vary only one parameter and keep the other two parameters fixed. We find that the expansion effects of QED corrections are going through decelerating growth with increasing $\mu$ and approximately linear growth with increasing $Q_m$, respectively. (Notice that we use a logarithmic coordinate for $\mu$ because of its wide range of variation) In contrast, the QED expansion effects remain basically invariant with increasing $a$, though a slight enhancement is still observable. Moreover, we find another relation with very high precision that $S/S_0\approx\left(\sigma_K+1\right)^2$ which reflects that black hole shadow is approximately a circle, even for relatively large $a$.

\subsection{The backreaction effect from QED correction}\label{backreaction}

In this and the next subsection, we will distinguish two different effects of 1-loop QED correction to the black hole shadow. The first one is the backreaction effect to the spacetime geometry, and the other one is the QED birefringence effect, leading to the QED enhanced gravitational traps as well as electromagnetic traps \cite{Novello:1999pg}, as explained in Section \ref{two concepts}.

\begin{figure}
    \centering
    \includegraphics[width=0.4\textwidth]{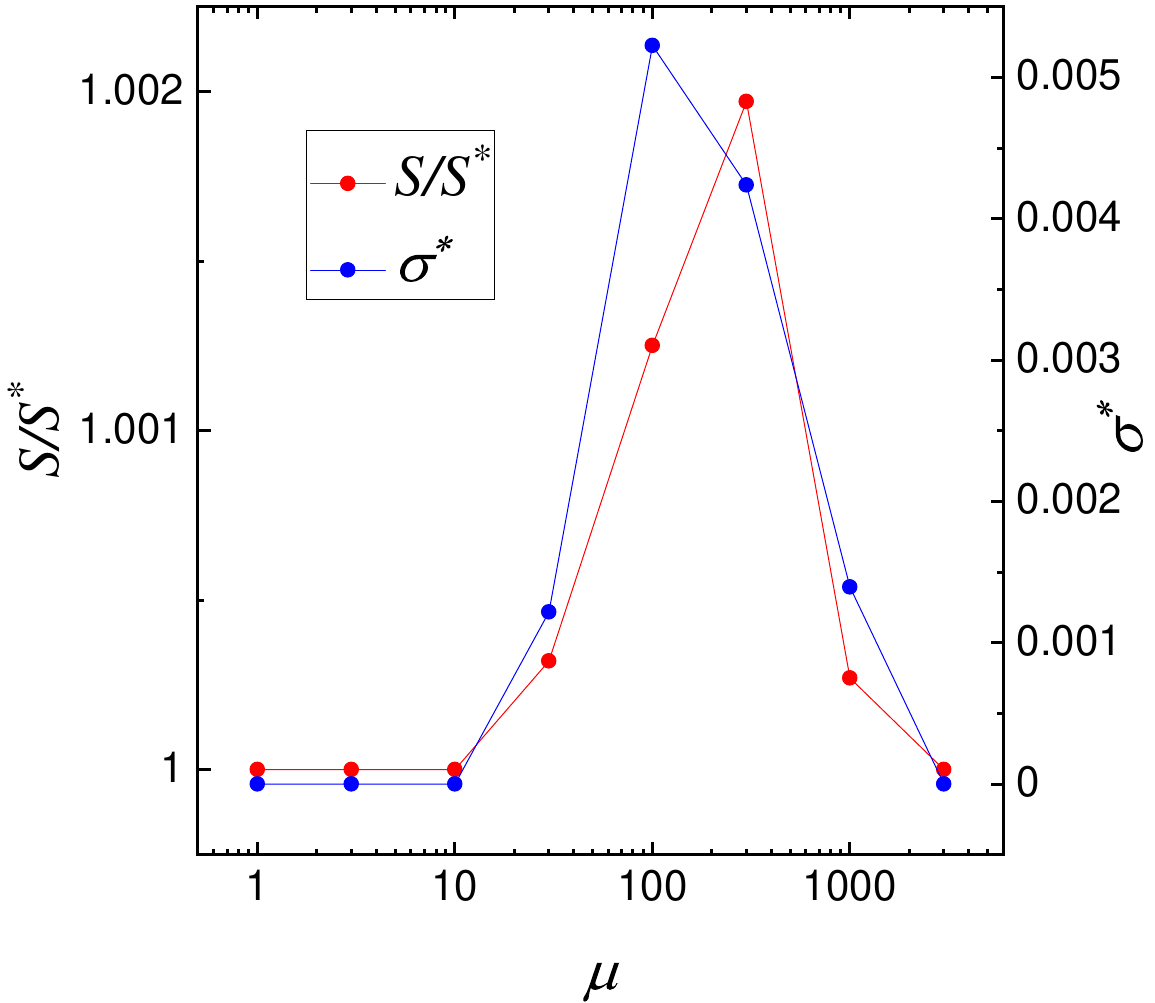}
    \caption{This figure demonstrates the contrast between the shadows with and without backreaction to the spacetime. We take $a=0.3$ and $Q_m=0.6$. $S$ refers to the area of shadows of given $\mu$ with backreaction to the background spacetime and $S^*$ refers to that without. $\sigma^*$ refers to the ``normalized" deviation between the corresponding shadows.}
    \label{fig:backreaction}
\end{figure}

Let us first investigate the contribution of the backreaction effects. We can do that simply by comparing the black hole shadows with and without the backreaction. Similar to Subsection \ref{geometrical analysis}, we define the geometrical parameters
\begin{equation}
    S^*=\frac{1}{2}\int_0^{2\pi}\Tilde{r}^*\left(\alpha\right)^2\mathrm{d}\alpha
\end{equation}
and
\begin{equation}
    \sigma^*\equiv\sqrt{\frac{1}{2\pi}\int_0^{2\pi}\left(\frac{\Tilde{r}\left(\alpha\right)-\Tilde{r}^*\left(\alpha\right)}{\Tilde{r}^*\left(\alpha\right)}\right)^2\mathrm{d}\alpha},
\end{equation}
where $\Tilde{r}^*\left(\alpha\right)$ refers to the shadow curve without considering the backreaction. The numerical results are shown in Fig. \ref{fig:backreaction}. As we can see from those results, the backreaction enlarges the shadow, which agrees with our discussion in Section \ref{section3}. However, the backreaction effect is rather tiny and can be neglected, as argued in \cite{Amaro:2023ull}.

Another interesting point is that the deviation between the shadows with and without background spacetime modification increases with $\mu$ at small coupling and then decreases. Their growth at small coupling is intuitively expected. When QED effects which expand the shadows get strong enough, the photons that go to infinity can't pass through small $r$ regions. As we can see, the background spacetime modification, i.e. the screening term shown in Section \ref{section3}, has a quartic attenuation with respect to $r$. Therefore, the modification becomes less important.

\subsection{ QED birefringence effects}\label{birefringence}

As explained in Section \ref{two concepts}, the QED birefringence effects make the trajectory of the photon timelike rather than null. Effectively, the photons move in a medium with a refractive index larger than $1$. If the electromagnetic field is local, the birefringence effects always enlarge the shadow, as shown in the case of a charged spinning black hole with QED corrections. 

\begin{figure}
    \centering
    \includegraphics[width=\linewidth]{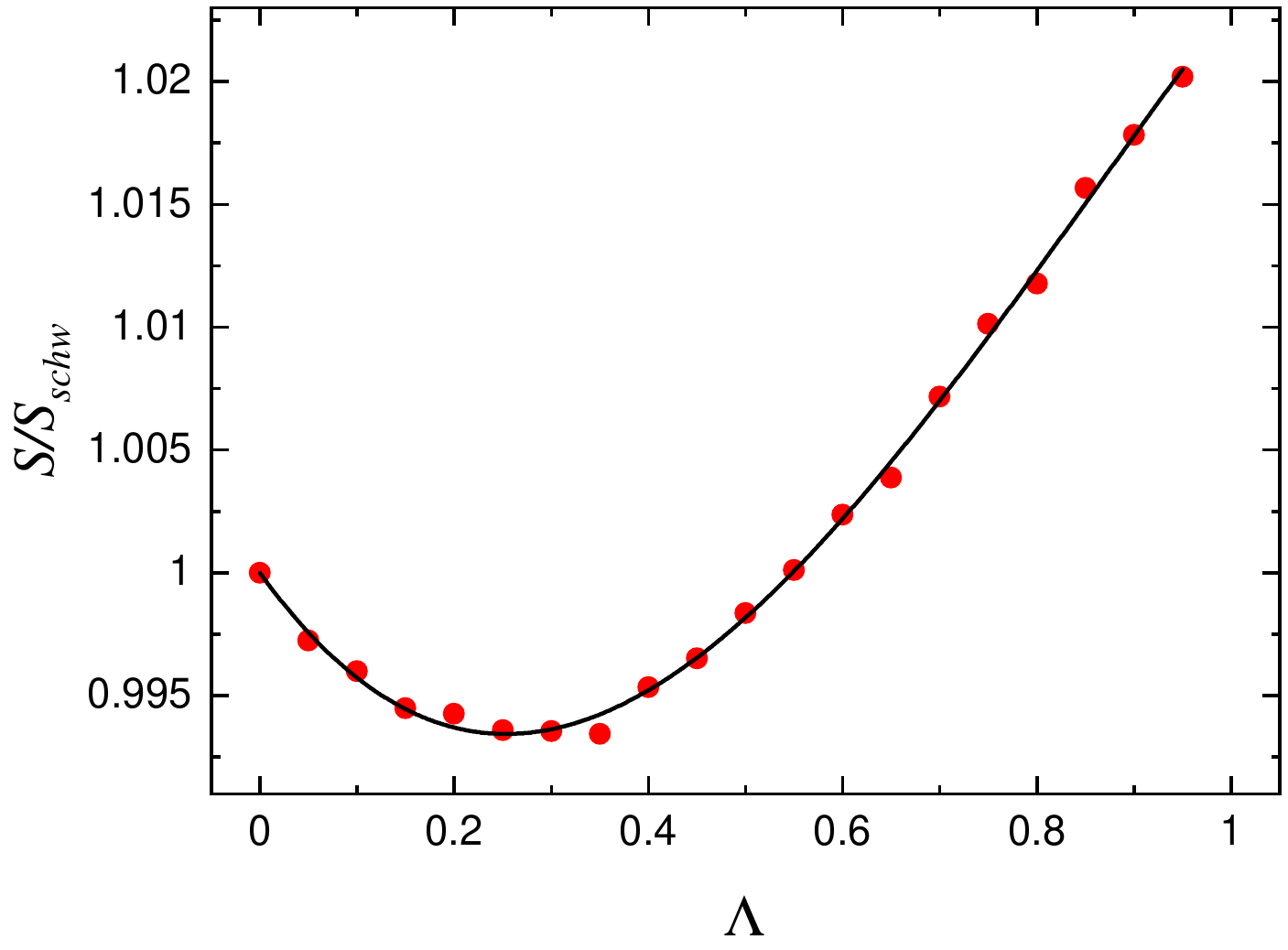}
    \caption{The variation of the area of the shadow with respect to $\Lambda$.}
    \label{fig:S-Lampda}
\end{figure}

However, the situation becomes subtle if the electromagnetic field is uniformly distributed. In the case studied in \cite{Hu:2020usx}, the black hole is immersed in a magnetic field extending uniformly to the infinity. As can be seen in Fig. \ref{uniform1}, the shadow is stretched along the magnetic field while is squeezed perpendicular to the magnetic field (Notice that the magnetic field is set to be horizontal in Fig. \ref{uniform1}). The areas of shadow at different $\Lambda$ are computed and plotted in Fig. \ref{fig:S-Lampda}. The area decreases a little with the increase of $\Lambda$ at first but eventually goes up rapidly at large $\Lambda$.

The perpendicular squeezing of the shadow as well as the initial decrease of the shadow area comes from the presence of the uniform magnetic field. This can be understood more clearly in flat spacetime, in which case the effective refractive index does not decay to 1 at infinity such that the light rays do not necessarily converge. It turns out that the uniform magnetic field tends to squeeze the shadow in a perpendicular direction. When $\lambda$ is small, the squeezing may induce the decrease of the shadow area. 

\section{Summary}\label{five sum}

In this paper, we conducted a study on the QED effects on the shadows of rotating black holes with magnetic charge. There are two distinct QED effects under consideration. One is the birefringence effect experienced by light rays in the strong electromagnetic field,  and the other is the extra distortion of the background spacetime due to backreaction. We considered both effects on the black hole images and found that the QED effects tend to enlarge the shadows of black holes.

In practice, we implemented the ray-tracing algorithm to numerically simulate the shadow of black holes with different parameters. We defined two geometric parameters, standard deviation of area and radius to characterize the expansion of black hole shadows quantitatively. The numerical results indicate that the QED-induced expansion of the shadows grows with QED coupling $\mu$ and magnetic charge $Q_m$ while having relatively little dependence on spin $a$.

At last, we analyzed the impact and contribution of different kinds of  QED effects. It turns out that both the QED birefringence effect and the backreaction effect tend to enlarge the shadow in our QED KN case. Also, our study demonstrated that the backreaction has a tiny influence on the black hole images and could be neglected safely, supporting the rationality in previous works \cite{Hu:2020usx,Zhong:2021mty}.

In \cite{Amaro:2023ull}, a similar topic has been addressed and it was found that the shadows of black holes would shrink under the QED effect, opposite to our conclusion. The discrepancy stems from the sign difference of $\lambda$, as we clarified at the end of Section \ref{two concepts}, even though the methods to read the shadow are different. In \cite{Amaro:2023ull}, in order to separate variables to get an analytical expression for black hole shadows, they took advantage of a bold approximation that $D_Q\left(r,\theta\right)\equiv Q^2/\left(D_c^2\Sigma^2\left(r,\theta\right)\right)=Q^2/[D_c^2\left(r^2+a^2\cos^2\left(\theta\right)\right)^2]$ is a quasi-constant on photon regions, where $D_c$ is a constant. In contrast, in this paper, we just numerically integrated Hamilton's Equations for better accuracy. Still, after correcting the sign, the two methods lead to similar pictures. 

\section*{Acknowledgments}

We would like to thank M.Y. Guo for the valuable discussions and his participation at the early stage of this work. S. Yuan would like to thank Y. Zimmermann for his help in providing the text of \cite{osti_4071071}. The work is partly supported by NSFC Grant No. 12275004.

\begin{appendix}

\section{The complete expression for effective metric tensor}\label{append}

Considering the most general covariant Lagrangian for the electromagnetic field with minimal coupling
\begin{equation}
    S=\int \sqrt{-g}\mathcal{L}\left(\mathcal{F},\mathcal{G}\right)\mathrm{d}^4x,
\end{equation}
where $S$ is the effective action of the electromagnetic field and $\mathcal{F}$ and $\mathcal{G}$ are the only two independent relativistic invariant and pseudo-invariant defined in Eq.(\ref{invariant}). Variation of the action $\frac{\delta S}{\delta A_\mu}=0$ gives the equation of motion
\begin{equation}
    \nabla_\nu(\mathcal{L}_\mathcal{F}F^{\mu\nu}+\mathcal{L}_\mathcal{G}(^*F)^{\mu\nu})=0,
\end{equation}
where $L_\mathcal{F}=\partial_\mathcal{F}\mathcal{L}\left(\mathcal{F},\mathcal{G}\right)$, etc.

Under the approximation of geometric optics, the photon's trajectory in nonlinear electrodynamics is a null geodesic of the effective metric $\Tilde{G}_\pm^{\mu\nu}$. The subscript represents the polarization of the photon. $\Tilde{G}_\pm^{\mu\nu}$ can be determined up to an arbitrary conformal factor, which doesn't change the null geodesics. Its complete expression is (see \cite{Novello:1999pg} for detailed discussions)
\begin{equation}
    \begin{split}
        \Tilde{G}_\pm^{\mu\nu}\propto&\left[\mathcal{L}_\mathcal{F}+\left(\mathcal{L}_{\mathcal{F}\mathcal{G}}+\Omega_\pm \mathcal{L}_{\mathcal{G}\mathcal{G}}\mathcal{G}\right)\right]g^{\mu\nu}\\
        &+4\left(\mathcal{L}_{\mathcal{F}\mathcal{F}}+\Omega_\pm \mathcal{L}_{\mathcal{F}\mathcal{G}}\right)F_\lambda^{\,\,\mu}F^{\lambda\nu},
    \end{split}
\end{equation}
where $\Omega_\pm$ are the two roots of the quadratic equation
\begin{equation}
    \Omega^2\Omega_1+\Omega\Omega_2+\Omega_3=0,
\end{equation}
with
\begin{align}
    \left\{\begin{aligned}
    \Omega_1=&-\mathcal{L}_\mathcal{F}\mathcal{L}_{\mathcal{F}\mathcal{G}}+2\mathcal{F}\mathcal{L}_{\mathcal{F}\mathcal{G}}\mathcal{L}_{\mathcal{G}\mathcal{G}}\\
    &+\mathcal{G}\left[\left(\mathcal{L}_{\mathcal{G}\mathcal{G}}\right)^2-\left(\mathcal{L}_{\mathcal{F}\mathcal{G}}\right)^2\right],\\
    \Omega_2=&\ \left(\mathcal{L}_\mathcal{F}+2\mathcal{G}\mathcal{L}_{\mathcal{F}\mathcal{G}}\right)\left(\mathcal{L}_{\mathcal{G}\mathcal{G}}-\mathcal{L}_{\mathcal{F}\mathcal{F}}\right)\\
    &+2\mathcal{F}\left[\mathcal{L}_{\mathcal{F}\mathcal{F}}\mathcal{L}_{\mathcal{G}\mathcal{G}}+\left(\mathcal{L}_{\mathcal{F}\mathcal{G}}\right)^2\right],\\
    \Omega_3=&\ \mathcal{L}_\mathcal{F}\mathcal{L}_{\mathcal{F}\mathcal{G}}+2\mathcal{F}\mathcal{L}_{\mathcal{F}\mathcal{F}}\mathcal{L}_{\mathcal{F}\mathcal{G}}\\
    &+\mathcal{G}\left[\left(\mathcal{L}_{\mathcal{F}\mathcal{G}}\right)^2-\left(\mathcal{L}_{\mathcal{F}\mathcal{F}}\right)^2\right].
    \end{aligned}\right.
\end{align}

\section{The Einstein Tensor and the Energy-Momentum Tensor}\label{deviation}

For a static black hole with either magnetic charge or electric charge only, up to the first-order QED corrections in the Euler-Heisenberg model, the Einstein field equations can be analytically solved to give the metric in Eq.(\ref{QED RN}). The four nonzero components of the Einstein tensor and the energy-momentum tensor can be calculated as
\begin{equation}
    \begin{cases}
        G_{tt}=8\pi T_{tt}=\frac{2m^\prime\left(r\right)}{r^2}[1-2m(r)/r],\\
        G_{rr}=8\pi T_{rr}=-\frac{2m^\prime\left(r\right)}{r^2}[1-2m(r)/r]^{-1},\\
        G_{\theta\theta}=8\pi T_{\theta\theta}=Q^2/r^2-3\mu Q^4/(2\pi r^6)\\
        G_{\phi\phi}=8\pi T_{\phi\phi}=G_{\theta\theta}\sin^2\theta,
    \end{cases}
\end{equation}
where $m(r)=1-Q^2/(2r)+\mu Q^4/(20\pi r^5)$ and $m^\prime(r)=Q^2/(2r^2)-\mu Q^4/(4\pi r^6)$.

In contrast, for a rotating black hole, the Einstein field equations can only be solved approximately to get the Gürses-Gürsey metric. There are six nonzero components in the Einstein tensor and the energy-momentum tensor, but only three of them are independent due to the following constraints \cite{Breton:2019arv}
\begin{equation}
    \begin{split}
        \frac{a\sin^2\theta G_{t\phi}+G_{\phi\phi}}{\sin^2\theta}&=\frac{r^2+a^2}{\rho^2}G_{\theta\theta},\\
        \frac{a^2\sin^4\theta G_{tt}-G_{\phi\phi}}{\sin^2\theta}&=-\frac{r^2+a^2+a^2\sin^2\theta}{\rho^2}G_{\theta\theta},
    \end{split}
\end{equation}
as well as the trivial $G_{t\phi}=G_{\phi t}$. The same constraints hold for the energy-momentum tensor as well.

The independent nonzero components of the Einstein tensor are given by \cite{Breton:2019arv}
\begin{equation}
    \begin{cases}
        G_{rr}=-2r^2m'(r)/(\rho^2\Delta)\\
        G_{\theta\theta}=-2m'(r)a^2\cos^2\theta/\rho^2-rm''(r)\\
        G_{t\phi}=\{2[(r^2+a^2)a^2\cos^2\theta-r^2\Delta]m'(r)\\
        \phantom{G_{t\phi}=}+(r^2+a^2)\rho^2rm''(r)\}a\sin^2\theta/\rho^6
    \end{cases},
\end{equation}
while those of the energy-momentum tensor are given by
\begin{equation}
    \begin{cases}
        8\pi T_{rr}=-\frac{Q^2}{\rho^2\Delta}(1-\frac{2\mu}{\pi}u)-\frac{2\mu}{\pi}(u^2+\frac{7}{4}v^2)g_{rr},\\
        8\pi T_{\theta\theta}=\frac{Q^2}{\rho^2}(1-\frac{2\mu}{\pi}u)-\frac{2\mu}{\pi}(u^2+\frac{7}{4}v^2)g_{\theta\theta},\\
        8\pi T_{t\phi}=-\frac{Q^2}{\rho^6}[\Delta+(r^2+a^2)]a\sin^2\theta(1-\frac{2\mu}{\pi}u),\\
        \phantom{8\pi T_{t\phi}=}-\frac{2\mu}{\pi}(u^2+\frac{7}{4}v^2)g_{t\phi},
    \end{cases}
\end{equation}
where
\begin{equation}
    \begin{cases}
        u\equiv\frac{Q^2}{2\rho^8}(\rho^4-8r^2a^2\cos^2\theta),\\v\equiv\frac{Q^2}{\rho^8}(r^2-a^2\cos^2\theta)2ar\cos\theta.
    \end{cases}
\end{equation}
Notably, these expressions apply for both the electrically and the magnetically charged case; $Q$ can either be the electric charge $Q_e$ or the magnetic magnetic charge $Q_m$.

The deviation from the Einstein field equations can be quantified by
\begin{equation}
    \delta_{\mu\nu}=\left|\frac{G_{\mu\nu}-8\pi T_{\mu\nu}}{G_{\mu\nu}}\right|.
\end{equation}
The explicit expression of $\delta_{\mu\nu}$ appears long and complicated; still, we might as well gain some intuitions from the asymptomatic behaviors. For the black hole parameters, with large black hole mass (small $\mu$), weak electromagnetic field (small $Q$), or slow black hole motion (small $a$), the deviation goes as
\begin{equation}
    \delta_{rr}\sim\delta_{\theta\theta}\sim\delta_{t\phi}\sim
    \begin{cases}
        O(\mu), &\mu\to0\\
        O(Q^2), &Q\to0\\
        O(a^2), &a\to0
    \end{cases},
\end{equation}
indicating the validity of the Gürses-Gürsey metric in said scenarios. In addition, with respect to the coordinates, as $\theta\to\frac{\pi}{2}$ or $r\to\infty$, we have the asymptomatic behaviors of
\begin{equation}
    \delta_{rr}\sim\delta_{\theta\theta}\sim\delta_{t\phi}\sim
    \begin{cases}
        O(\cos^2\theta), &\theta\to\frac{\pi}{2}\\
        O(r^{-6}), &r\to\infty
    \end{cases},
\end{equation}
indicating the additional effectiveness of the approximation at large distances and near the equatorial plane.

\section{Energy condition for the QED correction }\label{SEC}

In this appendix, we would like to study the energy conditions for the QED correction. We show that it violates both the strong-energy condition (SEC) and null-energy condition (NEC). As a result, the impact of the electromagnetic field on the gravitation gets attenuated. 

The strong energy condition requires that for any normalized timelike vector $t^\mu t_\mu=-1$, there is 
\begin{equation}
    R_{\mu\nu}t^\mu t^\nu\propto (T_{\mu\nu}-\frac{1}{2}Tg_{\mu\nu})t^\mu t^\nu=T_{\mu\nu}t^\mu t^\nu+\frac{1}{2}T\ge0,
\end{equation}
where $T\equiv T_\mu^\mu$ is the trace of the energy-momentum tensor. While the null energy condition requires that for any null vector $l^\mu l_\mu=0$, there is 
\begin{equation}
    R_{\mu\nu}l^\mu l^\nu\propto (T_{\mu\nu}-\frac{1}{2}Tg_{\mu\nu})l^\mu l^\nu=T_{\mu\nu}l^\mu l^\nu\ge0. 
\end{equation}

To begin with, it can be easily proved that the classical electromagnetic field (CEM) satisfies all four kinds of energy conditions. For example, as for the weak energy condition (WEC)
\begin{equation}
    T_{\mu\nu}t^\mu t^\nu\ge0,
\end{equation}
we can always find a coordinate frame where $\hat{t}=(1,0,0,0)$, and therefore
\begin{equation}
    \begin{split}
        T^{\text{CEM}}_{\mu\nu}t^\mu t^\nu=&\ \frac{1}{4\pi}\left[F_{\mu}^{\,\,\alpha}F_{\nu\alpha}-\frac{1}{4}\mathcal{F}g_{\mu\nu}\right]t^\mu t^\nu\\
        =&\ T^{\text{CEM}}_{00}=\frac{1}{8\pi}(\boldsymbol{B}^2+\boldsymbol{E}^2)\ge0.
    \end{split}
\end{equation}
This inequality may serve as a lemma,
\begin{equation}
    F_{\mu}^{\,\,\alpha}F_{\nu\alpha}t^\mu t^\nu\ge\frac{1}{4}\mathcal{F}g_{\mu\nu}t^\mu t^\nu=-\frac{1}{4}\mathcal{F}.
\end{equation}

Next, let us check the SEC for the energy-momentum tensor of the QED correction, presented in Eq.(\ref{DeltaT}),
\begin{equation}
    \begin{split}
        &4\pi(\Delta T_{\mu\nu}-\frac{1}{2}\Delta Tg_{\mu\nu})t^\mu t^\nu\\
        =&-\frac{\mu}{16\pi}\left[8\mathcal{F} F_{\mu}^{\,\,\alpha}F_{\nu\alpha}-\left(3\mathcal{F}^2+\frac{7}{4}\mathcal{G}^2\right)g_{\mu\nu}\right]t^\mu t^\nu\\
        &(\text{as long as } \mathcal{F}\ge0)\\
        \le&-\frac{\mu}{16\pi}\left[-\left(\mathcal{F}^2+\frac{7}{4}\mathcal{G}^2\right)g_{\mu\nu}\right]t^\mu t^\nu\\
        =&-\frac{\mu}{16\pi}\left(\mathcal{F}^2+\frac{7}{4}\mathcal{G}^2\right)\le0.
    \end{split}
\end{equation}
For a magnetically charged KN black hole, the condition $\mathcal{F}=2(\boldsymbol{B}^2-\boldsymbol{E}^2)\geq 0$ holds almost everywhere, except for the extremal spinning black hole case. In this work, we focus on the spinning black hole far from extremality, so $\mathcal{F}\geq 0$ always holds. Moreover, the larger the $\mu$ is, the more obvious the violation is. Similarly, we can show the violation of NEC for the QED correction. 

Furthermore, a similar argument can be made for the electrically charged KN black holes. The difference is that, in the electrically charged case, if we still use Maxwell variables ($F_{\mu\nu}$), we have to keep in mind that there are first-order QED corrections in them. In other words, though Eq.(\ref{electromagnetic energy-momentum tensor}) still holds, Eq.(\ref{DeltaT}) does not, since the proper variables are $P_{\mu\nu}$ as mentioned in Section \ref{section3} and $F_{\mu\nu}=F_{\mu\nu}(\mu)$. Anyway, the fact that the QED correction terms in the energy-momentum tensor tend to violate SEC and NEC still holds in the electrically charged case.

\end{appendix}

\bibliographystyle{utphys}

\bibliography{KNQED}

\end{document}